\documentclass[reprint,amsmath,amssymb,aps,pra]{revtex4-2}
\usepackage{pdfpages}
\usepackage{graphicx}
\usepackage{bm}
\usepackage{array}
\usepackage{enumitem}
\usepackage{MySwift}
\usepackage{booktabs}

\makeatletter
\AtBeginDocument{\let\LS@rot\@undefined}
\begin{document}


\title{Systemically Designed Degrees for Real-World Challenges: A case study on Physics curriculum design at Loughborough University}

\author{M.J.~Everitt}
 \email{m.j.everitt@physics.org}
\affiliation{%
Quantalytics, Loughborough, UK.
}
 \author{M.T.~Greenaway}
 \email{m.t.greenaway@lboro.ac.uk}
 \author{S.L.~Bugby}
 \author{S.N.A~Duffus}
\affiliation{%
Department of Physics, 
Loughborough University, 
Loughborough, 
LE11~3TU, 
UK.
}

\begin{abstract}
We present a ground-up redesign of the undergraduate physics degree at Loughborough University, driven by the principle of authenticity in academic and industrial practice. Departing from conventional incremental reforms, we adopt a systems-engineering approach to programme-level curriculum design, treating the degree as an integrated system with verifiable performance. This methodology aligns stakeholder-derived requirements with vertically-integrated threads in theory, computation, laboratory practice, and professional skills. We demonstrate that this approach enables students to achieve levels of disciplinary and cross-disciplinary competence beyond those typically expected at undergraduate level. Outcomes are supported by graduate destinations, enhanced student performance, and positive external evaluations, including national accreditation. Our results suggest that rigorous, system-level curriculum design can yield transformational gains in both capability and confidence.
\end{abstract}

\maketitle 
\tableofcontents

\section{Introduction}

It is an increasingly voiced view that the traditional physics degree is not adequately preparing  students for contemporary challenges and career paths \cite{Benson2023}. Across higher education in the UK, \emph{``Many employers still believe that graduates lack the ‘basic’ work-ready competencies that make up employability skills.''}~\cite{CMI2021}. Within the discipline of physics, although employers report that \emph{``physics graduates are better equipped than other graduates when it comes to tackling problems outside their field of expertise''} ~\cite{Sharma_2008}, in contrast capability gaps have been identified in areas such as computing, team-working, collaboration and communication, and problem solving (including experimental design and project planning) ~\cite{Sharma_2008}. These cross-cutting skills are only growing in importance to employers~\cite{dickerson2023analysis}. 
Aiming to address these concerns,  the Institute of Physics~\cite{IoP} and Quality Assurance Agency \cite{QAA3, QAA4} have recently taken decisive action by substantially revising their expectations of Physics degrees~\footnote{For transparency we note that one author (MJE) contributed to both ~\cite{IoP} and \cite{QAA3} as a parallel activity to that which we report here.}. This has changed the regulatory assurance frameworks for UK higher education, giving universities additional flexibility in curriculum design and delivery. In addition, across the UK HE landscape there is considerable concern regarding grade inflation \cite{Students2022} and a decline in the ability and knowledge of incoming students \cite{hawkes2000measuring}. Among physics educators, these factors have contributed to a widely voiced - though rarely evidenced - perception that the intellectual ambition of degree programmes has softened.

As with many continuing degree programmes, the physics programmes at Loughborough are iteratively updated year-on-year to incorporate new developments in physics and its applications, and in pedagogy. However, until 2019 it used the same fundamental structure as established several decades ago, this reflected a strong curriculum focus for degree accreditation - learning facts around physics phenomena but not developing skills and resilience in problem solving. 

Perspectives on this and related issues can be found in~\cite{Sands2019} who rightly states \emph{``The main outcome of the review of accreditation procedures is that the IOP Core of Physics is no longer a list
of topics that needs to be taught in the first two years.
Rather, it constitutes a set of themes that will run through
the entire degree and in a well-designed programme
students will develop their knowledge, understanding
and competence in applying that knowledge through-
out the duration of the degree. This allows for a much
more holistic approach to the design of the curriculum
and has the additional benefit that departments can be
much more distinctive in what they offer.''}

Over recent decades, UK physics curricula have been restructured so that advanced analytical topics such as a full treatment of Maxwell’s equations, elements of analytical mechanics, and Green’s-function methods are more often delivered later in programmes and or as options (or even not at all). Whether or not some of these changes indicate a lowering of expectations or simply a change in emphasis is not established in the literature.  What is apparent is these changes have, in our view,  happened organically and over time supported by institutional isomorphism rather than through coherent strategic interventions. These trends alone constitute a compelling case for a systematic, evidence-led review of physics curricula and learning outcomes.

Reflecting on the changing regulatory environment, the evolving career expectations of our graduates, and our commitment to maintaining and enhancing intellectual rigour, it was clear that innovation was required. 
We felt the time was right and the ambition was there to complete a from-the-ground-up programme redesign fit for the challenges of the 21st century.  A complete fresh start was an opportunity to reflect on and modernise our educational philosophy and teaching design principles alongside our content and delivery.  

Our ambition meant that we needed a coherent process to co-ordinate the redesign so we borrowed methodology from systems thinking and engineering. We have found this structured requirements-first approach valuable, and believe it would benefit the wider educational community.

Our new degrees opened in 2019 and have now graduated four cohorts of bachelors and three of integrated masters. 
We are confident that we have met our original aims, as evidenced against external benchmarks, and will use examples of our approach to illustrate the utility of systems engineering to curriculum design. 
Our final reflections consider the challenges we faced, and the challenges our sector has still to overcome.  

In this article we show how a systems-engineering approach to curriculum design can yield a physics degree that demonstrably enhances conceptual understanding, mathematical fluency, and graduate readiness, while remaining adaptable and scalable within diverse institutional contexts. 

\section{Systems Engineering in curriculum design}

Systems engineering is based on the principle that a solution is often greater, both in efficacy and complexity, than the sum of its parts.  
A physical analogy of systems-thinking is a circuit designed to be robust to electrostatic discharge; although the performance of individual components may be relevant, the robustness of the circuit can only truly be optimised by considering it as a whole. 
At its core, systems engineering is a requirements-driven methodology to design, implement and test solutions. 
Its fundamental philosophy is that you must understand \emph{what} you want to achieve, first and foremost, before considering \emph{how} you might achieve it.

Considering our degree programme holistically, as a complex arrangement of sub-components operating together to deliver set requirements in the context of institutional and sectoral constraint (along with departmental research interests in low TRL systems engineering), led us naturally to a systems engineering approach to curriculum design.

In education, instructional design (originally called instructional systems design) was borne from a desire for professional training in sectors with rigorously prescribed learning requirements. 
These sectors, led by the military, were also those which had already embraced systems engineering in other aspects of their work and it was natural that the same tools would be applied to designing training courses~\cite{molenda2009origins}. Instructional systems design became ubiquitous in professional settings but remained distinct from the curriculum design carried out in traditional educational establishments~\cite{rose2004instructional} with even proponents of instructional design such as Romiszowski~\cite{romiszowski2016designing} separating `instruction', which is goal orientated and pre-planned, from other types of learning (such as projects and research). 
More recently, and particularly post-COVID-19, the integration of instructional design in HE settings has greatly accelerated. 
However, its scope rarely extends to full programme curriculum design. 
In fact, the term instructional design in HE is most commonly used to describe the development of resources for short modules or training courses hosted online.

The use of Learning Objectives (LOs) is an example of requirements-driven design in HE. Although these are in common use, LOs are often criticised as overly prescriptive, particularly on a programme level~\cite{hussey2008learning}. 
The, also popular, `backwards design' approach of Wiggins and McTighe ~\cite{wiggins2005understanding} begins with a broad consideration of curriculum goals and expectations.  This is sufficiently broad that it is applicable to whole-programme level design but does not provide much guidance on the practical implementation of its stages. 

\begin{figure*}[t]
	\includegraphics[width=\textwidth]{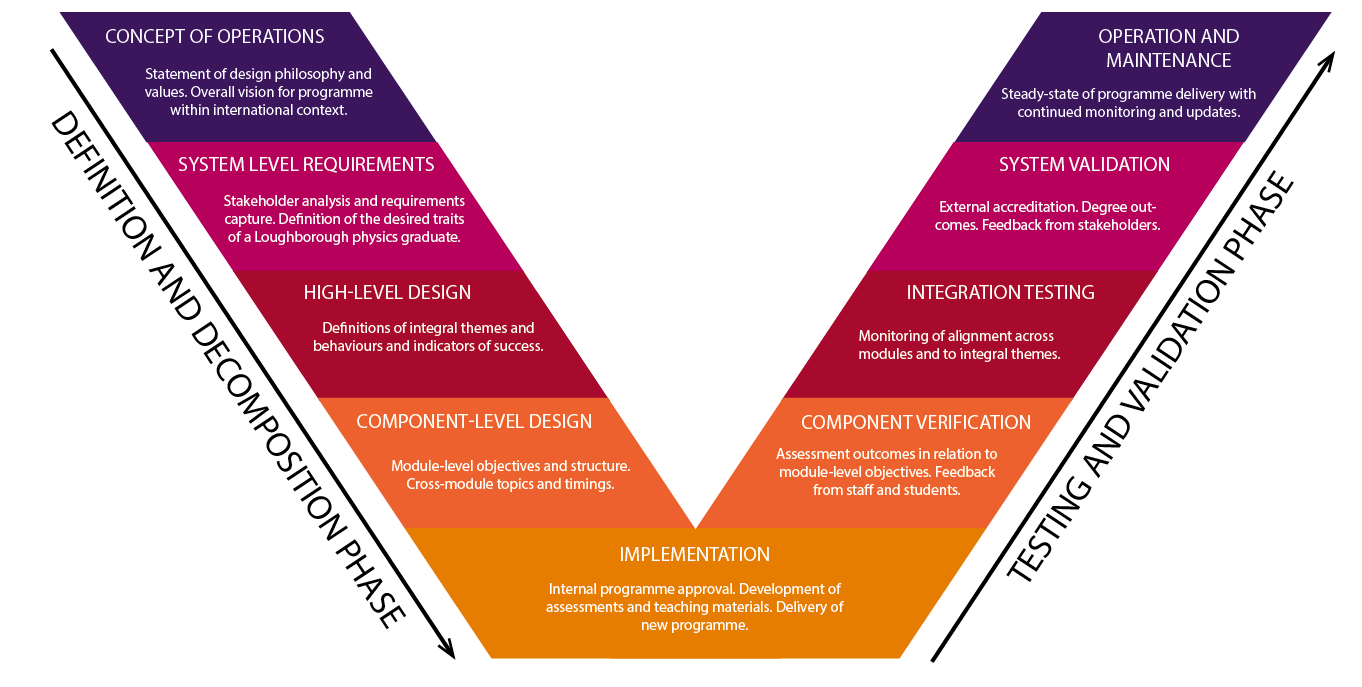}
	\centering
	\caption{Classic systems engineering V-diagram applied to curriculum development. Inherent in this model is a cyclical process of review and adaptation not shown for clarity.}
    \label{fig:V}
\end{figure*}

To bridge this gap we turned to the structure provided by systems engineering. Figure~\ref{fig:V} illustrates the classic 'V-diagram' of systems engineering, showing both the design and validation phases of a project, adapted for curriculum development. In subsequent sections we will expand on each aspect of this method and, through reference to our experiences with the Loughborough Physics degree, provide concrete examples of how it may be applied to curriculum design in HE. 
It should be noted that, although Figure~\ref{fig:V} and Section~\ref{sec:casestudydesign} imply a prescriptive and ordered process, in practice the approach is iterative, involving ongoing review, revision, and adaptation.

\section{The Loughborough physics degree: definition and decomposition phase}\label{sec:casestudydesign}

In this section we take each item of the design stage of the V-diagram, see Fig. \ref{fig:V} which is applied in classic systems engineering and show how it applied to the development of a physics programme.  

\subsection{Concept of operations}

We first make a  statement of our design philosophy and values to provide a vision of our programme.   Our overall aim was to develop \emph{physics degrees that provide graduates who industry and academia can trust to tackle formidable real world challenges}. With this in mind, and through discussions with staff and other stakeholders, we articulated our design philosophies:

\begin{enumerate}
    \item \textbf{Embed authenticity:} Physics is a practice-based discipline, the strength of physics comes from its application to the real world. Physics should matter, and our content and assessments should be drawn from the real world practice of physics (or physics-related activity) in academia and industry. As a corollary we inherit values of discovery, creativity, insight, veracity, honesty, exploration, diligence, perseverance, application of effort and attention to detail. This in turn leads to demands for depth, rigour and development of key skills; both academic and professional.
\end{enumerate}

\begin{enumerate}
    \setcounter{enumi}{1}
    \item \textbf{Embrace advanced unifying concepts early:} Some of the most universal concepts and approaches in modern physics (such as least action or Noether's theorem linking symmetries to conservation laws) are often seen as too complex for first year physics students. They are relegated to higher-level courses, and then applied retroactively to backfill topics already covered (e.g. mechanics, electromagnetism). We reject this as a necessity, and instead seek to build a strong and clear foundational understanding of these fundamental principles which subsequent theories are then built on 
    (an example illustrating how coding can be used to help students develop their understanding of challenging mathematical material from a physics perspective is given in Appendix~\ref{hchain} which demonstrates a numerical variational method to study of the elasticated hanging chain).  
\end{enumerate}

\begin{enumerate}
    \setcounter{enumi}{2}
    \item \textbf{Cultivate a growth mindset:} Loughborough University is known for its sporting excellence, and core to this is a culture of continuous improvement, where adversity is simply an opportunity for growth, which permeates throughout the institution. By carefully designing experiences that challenge students at key moments, we can create opportunities for growth, resilience, and skill development~\cite{Dweck2006,Duckworth2016}. Tasks that stretch students' ability or push them outside of their academic comfort zones are constructive, providing students with opportunities to learn, adapt, and grow~\cite{Yeager2012} and reinforcing coping strategies~\cite{Crum2013,McGonigal2015}.
\end{enumerate}

This articulation of our concept, allowed us to generate buy-in from internal and external stakeholders as we sought their contributions. It also served as the ultimate reference point, to be returned to when navigating difficult decisions in subsequent phases.

\subsection{System (degree) level requirements}

Following our systems engineering approach, once the overall aim and values of our programme redesign were established the next step was to consider specific requirements. It is important at this stage to focus on \emph{what} needs to be achieved, and to avoid any distraction from implementational practicalities which may pre-emptively narrow the design space. 

Our stakeholder analysis and requirements capture process consisted of desk research, to identify trends and drivers, and direct consultation with stakeholders (including staff, students, industry, professional bodies and national laboratories) to provide specific detail.  Primarily, the desk research focused on the experience of employers of physics graduates and physics graduates themselves. 
For example, the Institute of Physics had gathered feedback from employers indicating that physics graduates (sic~\cite{RH}):

\begin {itemize}
\item Have a good knowledge of physics but can be limited in what they can do with that knowledge; 
\item Their grasp of fundamental concepts can be weak;  
\item Communication, teamworking, personal, and professional skills can be poor; 
\item Can lack critical thinking, reasoning and complex problem solving skills; 
\item Need to be prepared to learn for themselves and reflect on their learning; Can find it difficult to describe what skills they have.
\end{itemize}

Stakeholder interviews provided richer detail, and allowed us to explore the underlying rationale behind requirements and to preserve the Loughborough context. Consultation with a diverse range of industry partners  made clear that the added value of the physicist to them was not simply the reductionist mindset of physics-thinking but also the ability to break down and solve \emph{very} hard problems.
During this period, there was also active contribution to QAA subject benchmark statement and IoP accreditation framework reviews to ensure a suitable regulatory framework was in place. 

Several cycles of planning, communication and evaluation yielded the implementation requirements of the degrees themselves. We initially used a trait-based approach; consolidating our findings into a statement of our aspirations for the Loughborough physics graduate (See Appendix \ref{LPUG}). 

From this process followed the programme-level learning outcomes \cite{progspec}, the primary requirements driving curriculum design. In this way, the systems engineering approach interlinks with more traditional curriculum design practices, with the programme learning outcomes acting as the top-level specifications from which all other elements are derived.

\subsection{High level design}

We now consolidate a high-level design for our programmes, with a focus on how the our system level requirements could be achieved within our concept of operation.   This meant the establishment of several themes which provide an architecture for the structure of the degree.  Here we define and expand each theme and reflect on what a successful outcome within that theme looks like.

\subsubsection*{Theme 1: Integration of theory, laboratory, computing \& mathematics}

A central design principle of the programme is to introduce unifying ideas in physics from the outset. For example, for classical analytical mechanics and electrodynamics -- typical first year material in a physics degree -- to be introduced from fundamental principles first requires exposure to Lagrangian and Hamiltonian dynamics, variational calculus, and the principle of least action. This presents a challenge; how can we equip students with the conceptual and mathematical depth required, despite their limited prior exposure (i.e. UK A-level physics students have typically not seen calculus deployed in the support of the discipline) without drastically altering the curriculum timeline?

To make this feasible, we restructured the curriculum to fully coordinate the delivery of mathematics, theoretical physics, laboratory skills, and computing. We focus on only one topic at a time, but \textbf{scaffold across delivery mechanisms}. This structure reduces redundancy, improves coherence, and encourages meaningful connections. See Section \ref{sec:weftandwarp} for a more detailed discussion of how this was implemented.

Success of this approach can be evaluated through student confidence in engaging with advanced concepts early in the programme, achievement in assessments, and evidence of conceptual transfer between domains. 

\subsubsection*{Theme 2: Challenge-based-learning}

The incorporation of challenge-based learning (CBL) across all years of study arises from the need in both academia and industry that physics graduates are able to solve very hard problems.

This approach, and indeed those that use real-world challenges as a defining characteristic, is not a new idea~\cite{CBL1} and neither is interdisciplinary content which is a common feature of complex projects~\cite{CBL2}. There is a range of different definitions of challenge based learning; because of our specific motivation, we include in our definition of challenge based learning that projects must not only be real world (such as unsolved problems specified by industry partners with real applications) but also sufficiently open that their complete solution is not guaranteed.

Importantly, CBL is not treated as an isolated module or capstone. Instead, it is scaffolded alongside technical content, with support from computational and laboratory instruction. For example, a first-year computational project requires numerical simulation of electromagnetic fields subject to temporal boundaries and time-domain metamaterials (See Appendix~\ref{AppC}) -- introducing students to uncertainty and partial knowledge, requiring them to apply and extend their understanding in unfamiliar contexts. 
As students progress through the degree, the scale and complexity of these challenges increase, culminating in final-year projects where even the requirements become the responsibility of the students (with input from external stakeholders).

Success may be evaluated from group project outputs, student self-efficacy in addressing open-ended problems, and feedback from academic and industry partners on students’ preparedness for real-world scientific work.

\subsubsection*{Theme 3: Substantial computational physics \label{sec:comp}}

Computational physics is incorporated into the degree as a core pillar rather than an adjunct skill, given a similar weighting as mathematics. This approach reflects both the realities of modern scientific research and the demands placed on graduates entering computationally intensive sectors such as data science, materials modelling, and imaging technologies. Computational physics supports theory (conceptual understanding as well as numerical experiments), challenge-based learning and experimental practice (simulations, data analysis, processing and control). 

To promote adaptability and professional readiness, students are exposed to multiple languages and tools throughout the programme, including Swift, C++, Julia, Python, MATLAB and COMSOL. The focus is not on specific language proficiency but transferable competence, training them to consider the best tool for a particular task and to be comfortable self-teaching languages or frameworks in the future. 
Given our principle of authenticity, we also explicitly teach the prinicples of good programming (e.g. lean, SOLID, TDD, FAIR~\cite{Wilkinson2016}, version control, validation, risk management). 

In short they need to begin to act like practitioners as outlined by the Alliance for Data Science Professionals \footnote{https://alliancefordatascienceprofessionals.co.uk/}.

Success can be evaluated from the quality and structure of student coding submissions, the application of computational techniques in non-computational modules (demonstrating transfer and integration of skills), and feedback from academic staff and industry partners on students’ readiness for computational challenges in research and professional settings.

\subsubsection*{Theme 4: Authenticity in laboratory practice}

Laboratory work in the redesigned programme is centred on authenticity, ownership, and professional relevance. Moving away from formulaic practicals with predetermined outcomes, our lab programme is structured to foster curiosity, experimental design, and critical analysis. 
Experiments are increasingly open-ended, span multiple sessions, and require students to record and communicate their work in a way that mirrors professional practice.

Assessment prioritises process over outcomes, with emphasis on critical thinking, iterative refinement, and evidence of problem-solving. 
Staff and postgraduate facilitators act as guides rather than instructors, encouraging students to independently troubleshoot and optimise their experiments. 
A representative challenge -- `design and build a polarimeter to investigate optical activity' illustrates the breadth of skills students must coordinate, including instrumentation design, control systems, data interpretation, and balancing competing constraints. Importantly the students are not dependent on existing laboratory equipment and engage in creating their own instrumentation for measurement, feedback, and control using, for example, Arduino controllers. Much like in a professional context, designs, calibration plans and uncertainty budgets are negotiated in-laboratory.

Key skills are embedded through structured support: planning workshops address time management, resource gathering, group collaboration, and risk mitigation; practical training includes CAD (e.g. SolidWorks), data analysis with OriginLab, and formal error propagation techniques. Students conduct full risk assessments -- including COSHH where appropriate -- and document their work in contemporaneous lab notebooks, stored in-lab to simulate professional practice.

Success is evaluated through student engagement with open-ended experimental design, the quality and completeness of lab records, and their ability to build, modify, and reflect on custom instrumentation in alignment with real-world scientific and engineering practices. In addition, we will assess the transfer and integration of experimental techniques into non-laboratory modules, the quality and structure of related coding and computational outputs, and feedback from academic staff and industry partners on students’ readiness for collaborative, interdisciplinary work in research and professional environments.

\subsubsection*{Theme 5: Group work and professional practice}

As mentioned in the computational physics theme, our drive for authenticity led to the teaching of coding practice not often covered in physics degrees (e.g. SOLID). 
In looking to integrate professional practice, we therefore not only look to disciplines directly within the domain of physics, but also to those careers where physicists often find themselves gainfully employed. 
Similarly professional practices such as as goal setting, systems engineering methods, and maintaining a laboratory logbook in-line with professional practice are integrated across computational, laboratory, and project modules. We consider group work a vital foundation of professional practice.

Effective group-work should result in an output that is greater than, or at the very least equal to, the sum of a coordinated but independently executed body of work. In order to experience a scenario where team-work adds value a degree of substantial complexity and difficulty needs to be in-place -- even in first year tasks, building as students' skills develop. 

For us, and in line with IoP accreditation requirements, teams must involve enough people that there is a complexity to organising each other and the work itself. In our work a team is an identifiable collaborative work-unit consisting of five or more people who are interdependent (so the whole achieves more than the sum of the parts) and who work towards a shared objectives with clearly assumed roles and responsibilities.

An example of the kind of challenge we expect our students to engage with is provided in Appendix \ref{Seabed}. Group work at this level of complexity cuts across all Loughborough Physics values. 
It is inherently authentic (particularly for projects proposed by external partners), requires a unification of advanced concepts, and can only be achieved if growth is embraced. 

Success is evaluated by outputs from group work. Capstone industrial projects are assessed with input from external partners against their professional expectations along with university staff for academic achievement.

\subsection{Component-level design: module objectives and cross module topics}\label{sec:weftandwarp}

\begin{figure}
    \centering
    \includegraphics[width=\columnwidth]{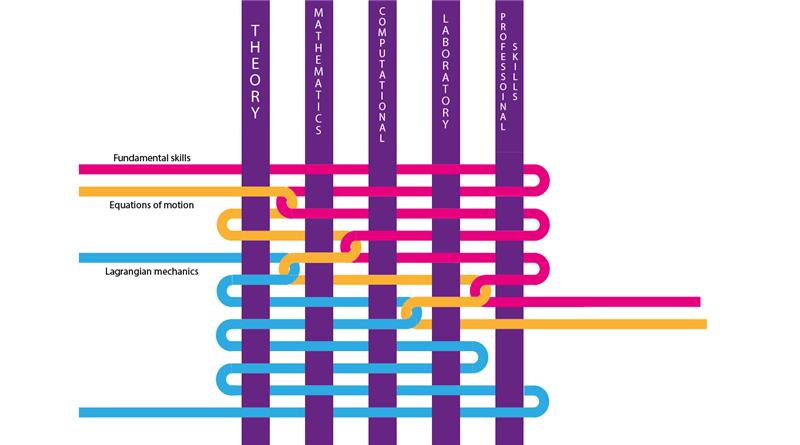}
    \caption{Illustration of how themes along and across the programme are interwoven to form the curriculum}
    \label{fig:weave}
\end{figure}

Putting the components of a programme together (modules - see \cite{progspec} for all module specifications) in order to achieve the above aims  and developmental themes is next part of the process. In much the same way as a cloth is made, we start with warp threads that run along the length of the programme. Our five broad threads are theory, laboratory, mathematics, computing, and professional skills, each which may contain sub-strands e.g. systems engineering and group work. To turn warp threads into cloth, we weave across the weft threads, using themes or topics to connect learning across the programme. Figure~\ref{fig:weave} illustrates this curriculum tapestry approach, but it is perhaps best explained through examples.

\begin{itemize}[label={}]
\item \textbf{Along-programme (warp thread) example: Systems engineering}
As part of our professional skills offer, training in systems engineering runs throughout our curriculum. In their first year, students are introduced to software requirements in their computational physics modules. In their second year laboratory module, students use a system requirements template to formalise performance requirements for instrumentation they design, build, and test. In the third-year group project, they collaborate with an industrial partner to agree on a user requirement specification, from which they derive a system requirement specification (SysRS) along with several systems engineering artifacts (e.g. Function Means Analysis) and a compliance matrix. The final assessment in this project evaluates technical performance based on how well these requirements are met, introducing students to verification and validation (V\&V) practices. 
Repeated engagement with a skill (i.e a spiral curriculum ~\cite{bruner1960process}) allows students to consolidate understanding, deepen conceptual connections, and transfer knowledge to new contexts.
\item \textbf{Across-programme (weft thread) example: Waves}
This approach is best illustrated by an example from the first-year, second-semester curriculum, where students undertake a comprehensive study of oscillation and wave phenomena. In their Core Physics module, students are introduced to the theory of electromagnetic field propagation. In mathematics, they study partial differential equations and boundary value problems. Computational physics brings this content together, implementing numerical algorithms (i.e. finite-difference time-domain) to explore waveguide design as part of a group project spanning several weeks. In the laboratory, students compare and contrast light and sound waves and experiment with mechanical waveguides. 
By bridging theory and practice and encountering the same topic in different contexts, students gain fresh insight and are more likely to successfully transfer their knowledge to new and complex problems \cite{billing2007teaching}.
\end{itemize}

This level of interconnectivity of topics across modules necessitated a strict 'top down' assignment both of module ILOs and the timetable of content delivery. In this respect, the use of a systems approach to curriculum design diverges somewhat from typical university practice i.e. achieving systemic optimisation often requires compromises at the granular level of curriculum delivery. This means that we will sometimes need to compromise on the best delivery of a given topic to give students the best learning experience over their entire programme of study (not just within years but also between those years).

\subsection{Implementation}

Planning for the new degree programme began in 2014, with curriculum plans assessed and approved through standard university quality assurance governance mechanisms (e.g. yearly programme review board).  The development of new programmes undergo additional scrutiny by the university's learning and teaching committee, chaired by the pro-vice chancellor for teaching and are ultimately approved the senate.   The first students entered the programme in October 2019. As expected with the introduction of a new programme, a number of modifications to the original plans were necessary during the first few years (some of which are explored in greater detail in the following section). Initially, the new programme was delivered in parallel with the previous one to accommodate students returning from placements or study interruptions who needed to complete the original curriculum.

\section{The Loughborough Physics degree: a case study in systematic curriculum verification and validation}\label{sec:casestudyeval}

The first graduates of the new programme were July 2022 (for BSc students) and July 2023 (for MPhys students). We have now graduated three or more complete cohorts from the new programme and hence it is an appropriate time to reflect on its successes and challenges via the testing and validation phase. Verification and validation (V\&V) is as important to the effectiveness of a systems engineering approach as the initial design stage. As with design, V\&V occurs at increasing levels of complexity throughout the programme implementation. For effective management, it is essential that V\&V processes are not siloed from design activities but instead allow iteration and continuous improvement. In this section we provide examples of programme refinements that illustrate this cyclical and responsive approach to curriculum development.

Assessment of the programme’s effectiveness has been undertaken through multiple lenses, guided by Brookfield’s reflective framework. In the early stages, evaluation necessarily relied on reflective practice, drawing heavily on the autobiographical lens of the senior teaching team, colleagues and student perspectives. As delivery progressed and the programme could be viewed as a complete system, external evaluations were incorporated. The new programme was accredited by the IOP in 2023 and extracts from the accreditation report have been integrated throughout to underscore our reflections.  Quantitative metrics are also considered however the inherent delay and high-level nature of such data mean they do not fully replace the value of timely, subjective insights.

\subsection{Component verification}

Evaluation on a module-level did not deviate too much from standard practice, with staff reflections, student feedback, and module results incorporated into the annual module update process. 

Early success was seen with challenge-based learning and authenticity in lab practice with staff observing a greater confidence in students to attempt problems where they didn't have a recipe to follow. This was also noted in the IOP accreditation report: \emph{``The open and exploratory nature of the experimental work impressed the panel. The students appreciated the value of this approach. The students reported that the laboratories are not scripted but that they can follow what to do without scripted instructions. They noted it is important to make mistakes and learn from them. Initially some students could find it quite difficult to be left to their own devices but came to appreciate being able to plan and execute the experiment for themselves."}

\begin{itemize}[label={}]
\item \textbf{Example refinement: open-ended lab activities}

The first year laboratory module is an example of module-level adaptations that were needed. The design of open-ended laboratory experiments, while innovative, sometimes ventured too far into a constructivist approach. Although such methods foster independence and creativity, some students found the lack of structured guidance too challenging. 
In order to better manage the transition we introduced a competency framework which provided a degree of structure whilst being sympathetic with the challenge based approach. This helped  students gain the confidence needed to engage in fully open-ended investigations, better supporting diverse learning needs and enhancing the overall learning experience.

\item \textbf{Example refinement: Scope of Coding and Computational Content}

One area of reflection concerns the volume and depth of coding exercises incorporated into the programme. The original ambition of covering all coding paradigms including functional and constraint based was too great. Moderating our delivery in the first year to Swift (for modelling and simulation) and C (for Arduino) and only four additional languages in the second year of study (e.g. for reproducing recently published numerical results). This approach helped the students focus their efforts on tasks that they perceived to be achievable and allowed for a depth of understanding of the languages that would not have been facilitated in the original approach. The sense of achievement obtained, and confidence gained, by reproducing recently published material in individual coursework provided a strong foundation for future work of an open nature.

\item \textbf{Example refinement: rebalancing between outcomes and process in group projects}

Within laboratory modules, we made a deliberate choice to reward process over outcomes. Given the open-ended nature of much of the lab work, where it is impractical to pre-evaluate every possible student approach, this was essential to ensure that creative or higher-risk, higher-reward strategies were not penalised, provided they adhered to sound scientific practice.

Applying the same process-over-outcome approach to third-year, industry-proposed projects initially led to some unintended consequences. Given the extended duration and professional context of these projects, it became clear that a reasonable expectation of tangible outputs was appropriate. As a result, the assessment weighting was rebalanced to place greater emphasis on deliverables. Students are still encouraged to propose stretch goals or explore more disruptive solutions, but these are now framed within their own planning and risk mitigation strategies and are no longer sufficient justification, on their own, for not delivering core project outcomes.

\end{itemize}

\subsection{Integration tests}

With the curriculum tapestry approach, all modules were strongly interconnected, despite being delivered by different module leaders and, in some cases, across different departments. This necessitated close collaboration among staff, but also required a co-creative approach to programme development. Only students had first-hand experience of the flow of topics across modules and the overall coherence of the curriculum, so their reflections were a vital component in V\&V. Feedback from both staff and students was positive, with the integration of topics across modules seen to accelerate students' understanding of complex topics. From the IOP accreditation report: \emph{``The students indicated that the interlinking of the physics, mathematics and computational content is a good structure as it helps them see the connections between different areas of physics."}

Another positive that emerged was the enhanced level of engagement from students on the programme. We attribute this partly to the authenticity of learning activities, and also to the strong departmental community which had long been a strength at Loughborough but was strengthened by the style of workshops and drop-in support sessions introduced in the new programme. 
As noted by the IOP accreditation report: \emph{``When the students were asked to describe the environment and culture in the department the first word that came to mind was inclusive. This was very pleasing to hear."} This proved particular benefit during 2020-2021 where lockdowns associated with the COVID-19 pandemic required a dramatic shift in delivery mechanisms -- we were lucky to be able to run online sessions with substantial interaction from students.

Staff-student liaison committee  meetings were able to take an overarching view of the programme (distinct from the module-level feedback delivered as standard) and identify pinch points or areas of development. 
Some of these co-created developments were straightforward to implement e.g. an alignment of terminology across modules taught by different departments, whereas others challenged staff assumptions and led to more fundamental shifts in our approach. 

\begin{itemize}[label={}]

\item \textbf{Example refinement: balancing challenge and support}

A key learning from staff–student co-creation was the contrast in perceptions between year groups. While second-year students often found full-day labs and escalating project demands overwhelming, final-year and post-placement students consistently reflected on the long-term value of these experiences. They highlighted how the structured challenges had built stamina, independence, and transferable skills such as time management and resilience, benefits that only became clear through later application.

This finding reinforced our pedagogical intention: to design the programme around a deliberate escalation of challenge, aligned with our values of authenticity and growth mindset. In their first year, students complete a short group project with a defined outcome; in the second year, they design and conduct their own experiment over a semester; and in the final year, they independently manage a year-long project with an external stakeholder. A similar pattern of escalation was embedded throughout the programme, not only in the increasing depth of physics content but also in the development of transferable skill development, such as clustering deadlines to promote forward planning.

As students navigate through these structured challenges, it is our intention that they learn to view demanding requirements as an opportunity for growth rather than a hindrance. 
This approach not only prepares them for academic success but also equips them with essential skills for their personal and professional lives. 
Ultimately, the goal is to foster a community of learners who are empowered, resilient, and ready to tackle the complexities of the world with confidence.

However, the balance was not always right. Some students, particularly in second year, experienced the workload as unmanageable. In response, we acted to maintain appropriate challenge while adding support: reducing unnecessary barriers (e.g. the number of programming languages), introducing more scaffolding and resources, and offering one-on-one coaching where needed. We also created more opportunities for cross-year dialogue, so students could share strategies and see the longer-term value earlier.

Communicating the purpose behind these design choices, and sharing retrospective feedback from previous cohorts, further helped students contextualise their experience. As noted in the IOP accreditation report of our programme, \textit{“The appreciation the students had for the skills they were developing through their programme was impressive.}”

\end{itemize}

\subsection{System validation}

We were only able to begin evaluating the programme in its entirety once the first cohort reached the end of their studies. In practice, the expected need for adjustments during the initial delivery, combined with the disruption caused by the COVID-19 pandemic, meant that it took several additional years before we could confidently assess the programme in its intended, stable form.

External accreditation by the IOP provided an important opportunity to reflect on the implementation of the new programme and to ensure that its quality was subject to rigorous and independent evaluation. 
The accreditation report was supportive of all changes made in the new programme particularly its authenticity in teaching and assessment, approach to laboratory work, substantial computational components, and the level of student independence fostered by the programme. From an accreditation perspective, the new programme was fully validated.

\emph{``The panel highlighted the redesign of the programmes as good practice, noting that it had been a very ambitious undertaking that had been executed well. The revised programme sounded fresh, distinctive, innovative, and exciting....The revised degree structure and programme of study is impressive. The programme is specifically designed to encourage students to become independent learners and evidence of this being successful was picked up during discussions with the students. The students met by the panel were certainly enthused by their studies and communicated their enthusiasm to the panel."}

Our initial aim was to design a physics degree that equips students with the knowledge and skills needed to thrive in an increasingly complex and dynamic world. As such, the most meaningful measure of success is the success of our graduates.

One early indicator of student success is their performance in third year industry proposed projects. These are almost entirely self-managed by students, with staff mentors meeting each group for less than 10 hours over the full year and acting in an advisory role (in terms of processes, or avenues to explore) rather than subject specialists. 
At the time of writing, we have run 3 cycles on industry proposed projects with many of our external partners choosing to reprise their role across multiple years.

Feedback from external partners is strongly positive. For example, they were asked to evaluate students against the expected performance for a recent graduate recruit. 
Across 4 indicators (communication, work ethic, independence, problem solving) and 8 project groups in the 2024/25 cohort -- students at least `met professional expectations' in 89\% of cases, with 48\% 'exceeding professional expectations' and 17\% 'substantially exceeding professional expectations'. 
Several of our students have gone on to employment in our external partners which is another vote of confidence in their ability.

Both the National Student Survey (NSS) and the Graduate Outcomes Survey (GOS) have undergone changes over the same period as the implementation of our revised programme. The Covid-19 pandemic and ongoing effects have undoubtedly changed the higher education landscape and graduate opportunities. Given this, a direct statistical comparison would not be appropriate however we were able to use NSS and GOS results internally -- from a qualitative perspective alongside free text responses -- to validate that the programme had met our aims. We also note that the implementation of the new programme has coincided with a rise in rankings in UK-based university league tables. Although, again, these are not statistically rigorous, they do provide some indication that the new programme is making a difference.

\subsection{Operation and maintenance}

The new programme has now reached steady state. We will continue to make updates to the curriculum and delivery based on developments in physics and pedagogy as part of the University's annual update cycle, but these modifications now tend to manifest as minor adjustments. This suggests a convergence towards a curriculum that effectively balances depth and breadth, preparing students comprehensively for future academic and professional pursuits.

Despite this, the systems approach still has value when looking forward to future challenges. 
For example, the rise of generative artificial intelligence -- with models such as ChatGPT or Microsoft Copilot becoming ubiquitous -- is creating both challenges and opportunities across the sector. 
At Loughborough, our prime value of authenticity tells us that, as the use generative-AI in the workforce is only increasing, we should support our students to use it in an authentic and appropriate way. 
The integration of generative-AI use as a professional skill flows logically from this, and the open-ended and challenging group work -- too complex to be carried out by gen-AI alone -- allows students to use a now industry-standard technique while demonstrating their problem solving skills and critical-thinking skills.  

\section{Reflections on Systems Engineering in curriculum design}

Systems engineering methods were integral to the programme’s design but, in hindsight, could have been used to even greater effect. 

While we adopted a requirements-driven approach, we did not initially make full and explicit use of a holistic requirements model. 
Our design-for-traits methodology led to the identification of both operational requirements (defining the major purpose) and functional requirements (specifying what must be achieved, rather than how). Our treatment of non-functional requirements was less formally articulated than it should have been and was not explicitly categorized into performance, system, or implementation requirements. 
As a result, we may have made the common process error, often seen among physicists, of moving too rapidly toward solutions without fully exploring the broader design space. 

A clearer and more systemic articulation  and use of requirements modelling would have enabled us to leverage tools such as Systemic Textual Analysis or Function Means Analysis. These methods would have facilitated a more systematic exploration of the trade-space of design options, helping us evaluate alternative ways of meeting our programme objectives. Additionally, incorporating tools such as Functional Failure Modes and Effects Analysis (FFMEA) from the outset could have helped pre-empt issues that were only later identified and addressed with hindsight. For example the function of delivering a scaffolded curriculum is vulnerable to unbalanced disruptions to the teaching schedule and student absence. Better articulating the effects of such such failure modes would be an example of reliability engineering of the programme. 

While it is not likely to Covid itself could be planned for, it is certainly the case that access to laboratory facilities could at any stage be compromised. Preplanning take-home laboratory experiments  in advance of issues arising would have been a likely outcome of performing an  FFMEA. In practice staff did respond, swiftly and creatively, with take-home experiments based on Arduinos for measurement and control.  Serendipitously, our original plan -- using Arduino-based instrumentation and measurement that students code themselves to build in-laboratory independence and resilience -- also strengthened the delivery resilience of our programmes.

A more rigorous analytical approach would also have enhanced our preparation for accreditation, easing the burden of  ensuring that the programme met all necessary criteria with each and every change. Similarly, formalised success metrics devised at the start of planning, particularly the collection of baseline data, would have made it easier to assess the success of the changes made.

There was a missed opportunity as explicit documentation and management of systems engineering artefacts were not as comprehensive as they could have been. 
A more formalised record-keeping system would have enhanced the traceability of design decisions and provided a clearer rationale for implementation choices. 

The creation and management of systems engineering artefacts serve not only to clarify objectives and improve implementation but also to establish a structured record of knowledge. 
These artefacts document both intended outcomes and the rationale behind specific design decisions, providing long-term traceability. 
In some instances, we have lost institutional knowledge regarding the full justification for certain decisions, highlighting the importance of structured documentation. 
Better management of these artefacts would have improved the traceability of our design and implementation processes, ensuring that institutional memory is preserved.

While such issues may not be critical during steady-state operation of the programme, historical knowledge of prior design choices is invaluable when evaluating and evolving the curriculum. 
Systems engineering artefacts, therefore, provide lasting value beyond the initial implementation phase, supporting continuous improvement and informed decision-making in curriculum development.

\section{Concluding remarks}

It is important to emphasise that we are not advocating that all physics degrees copy what we have done. What we do advocate though is that a \emph{design-for-traits}/systems thinking approach is adopted based on departmental values and ambition. 

Our reimagining of the physics degree was driven by a mixture of dis-satisfaction with the status quo together with a recognition that traditional curricula often fail to fully equip graduates with the depth of knowledge, adaptability, and professional competencies required in the contemporary work-place. By adopting a systemic approach to programme design, drawing from principles of systems engineering, challenge-based learning, and professional authenticity, we have formulated a holistic approach to academic programme development that combines theoretical rigour with applied problem-solving, and professional development.

A key achievement of our redesign has been the integration of learning experiences that reflect the realities of academic and industrial physics. Our programme places significant emphasis on scaffolding foundational knowledge in mathematics, computation, and experimental methods while promoting early exposure to complex, real-world problems. 
By embedding computational physics, systems thinking, and authentic laboratory work throughout the curriculum, students are encouraged to engage with the subject matter in a manner that is reflective of professional practice. 
The deliberate introduction of structured challenges further inculcates resilience, critical thinking, and adaptability.

One of the most notable outcomes of our curricular transformation has been its validation through external accreditation, employer feedback, and student engagement. 
The recognition by the Institute of Physics, alongside positive endorsements from industry partners, suggests that the programme is succeeding in producing graduates who are not only technically proficient but also highly employable and prepared for interdisciplinary problem-solving. 
Moreover, student feedback has revealed a trajectory of initial challenge and adjustment followed by retrospective appreciation of the rigorous and integrated nature of their training, evidence that the growth mindset cultivated within the programme is yielding long-term benefits in student preparedness.

While the programme represents a significant departure from conventional physics degree structures, it also serves as a model for how systemic, values-driven curriculum design can be employed to modernise education. 
The approach taken here demonstrates that the design of an effective degree is not merely an exercise in content optimisation but rather an exercise in aligning educational philosophy with desired graduate attributes. 
The success of this initiative suggests that curriculum design should be approached with the same methodological rigour as scientific inquiry, iterative, evidence-based, and responsive to the evolving demands of both academia and industry.
Navigating the landscape of systems thinking is in and of itself a challenge. For this reason we have provided an additional selection of standard systems engineering tools, their place in the V-diagram, and their function in Appendix~\ref{toolstable}. Further details are beyond the scope of this paper.

As we continue to refine and adapt this programme, we acknowledge that educational design is an ongoing process rather than a static achievement. 
The lessons learned through this undertaking reinforce the necessity of continual curriculum evaluation. 
Ultimately, by cultivating a learning environment that mirrors the complexity and demands of real-world physics, we believe that this programme is not only redefining physics education at Loughborough University but also offering valuable insights for broader educational reform.

\bibliography{apssamp}

\section{Author Contributions}


\noindent Mark Everitt: Conceptualization, Methodology, Validation, Writing -- Original Draft. Mark Greenaway: Methodology, Validation, Writing -- Review \& Editing. Sarah Bugby: Methodology, Validation, Writing -- Review \& Editing. Stephen Duffus: Methodology, Validation, Writing -- Review \& Editing\\

MJE was Director of Studies (DoS) in Physics at Loughborough University until 2024 and led the development of the programmes reported here.

\section{Acknowledgments}
MJE would like to thank the IoP for the opportunity to be involved in its Quantum Curriculum Project, which played a pivotal motivational role in initiating the curriculum redesign process. MJE would also like to thank the QAA and IoP for the opportunity to participate in subject benchmark and accreditation review processes respectively. The insights gained from these experiences, along with the ability to contribute to their outcomes, were instrumental in the successful redesign of our programmes. We are particularly thankful to all our current and previous staff and students for their input. Externally we had significant support from examiners, the QAA subject benchmark review group, the IoP accreditation review group and industry stakeholders.

ChatGPT was used the sole purpose of refining and clarifying text.

\appendix


\clearpage
\section{Code listings\label{hchain}}
Code is in example for the way that elasticated hanging chain can be used to teach the variation method by minimisation of potential energy.
Note that this code is developed in an interactive lecture from scratch and, because of this, differs  every year of delivery. We start by breaking the problem down into its conceptual component parts.

In the spirit of test driven development or at least using leads driven development a main driver routine is outlined first with data types and methods needed to make the algorithm work assumed to exist.
This helps us understand the type of structures and methods that will be needed to implement the algorithm.
The structure ChainElement is usually the first structure to be written as it is clear that a chain comprises many elements and we need this in place to build a chain.
In the first instance only the initialiser is provided and additional methods added on an as needed basis in the next phase will begin to define the Chain structure.
Chain is then built up bit at a time based on knowing the Physics ingredients that we will need. 
The initialiser simply sets up a chain of elements distributed evenly in a straight line between the end points of a chain (we can discuss clean coding as to e.g.  whether or not the function heightY is defined in the correct place).
The first method to be added due to its simplicity is the gravitational potential energy.
The more complicated methods of built-up through interactive discussion and exploration (making mistakes and fixing them).
Idea is to get reasonably clear and correct code but not at this stage focus on performance of the algorithm.
We discuss the clarity of the code and are critical of where it is not self documenting.
Note that we try to comment intent and not function. Also note that plotting the data is done with a separate package to try and emphasise separation of concerns.

\onecolumngrid

\lstinputlisting[language=Swift]{hangingchain.swift}
\clearpage

\includepdf[pages={1},frame,scale=.95,pagecommand=\section{The Loughborough Physics Graduate\label{LPUG}}]{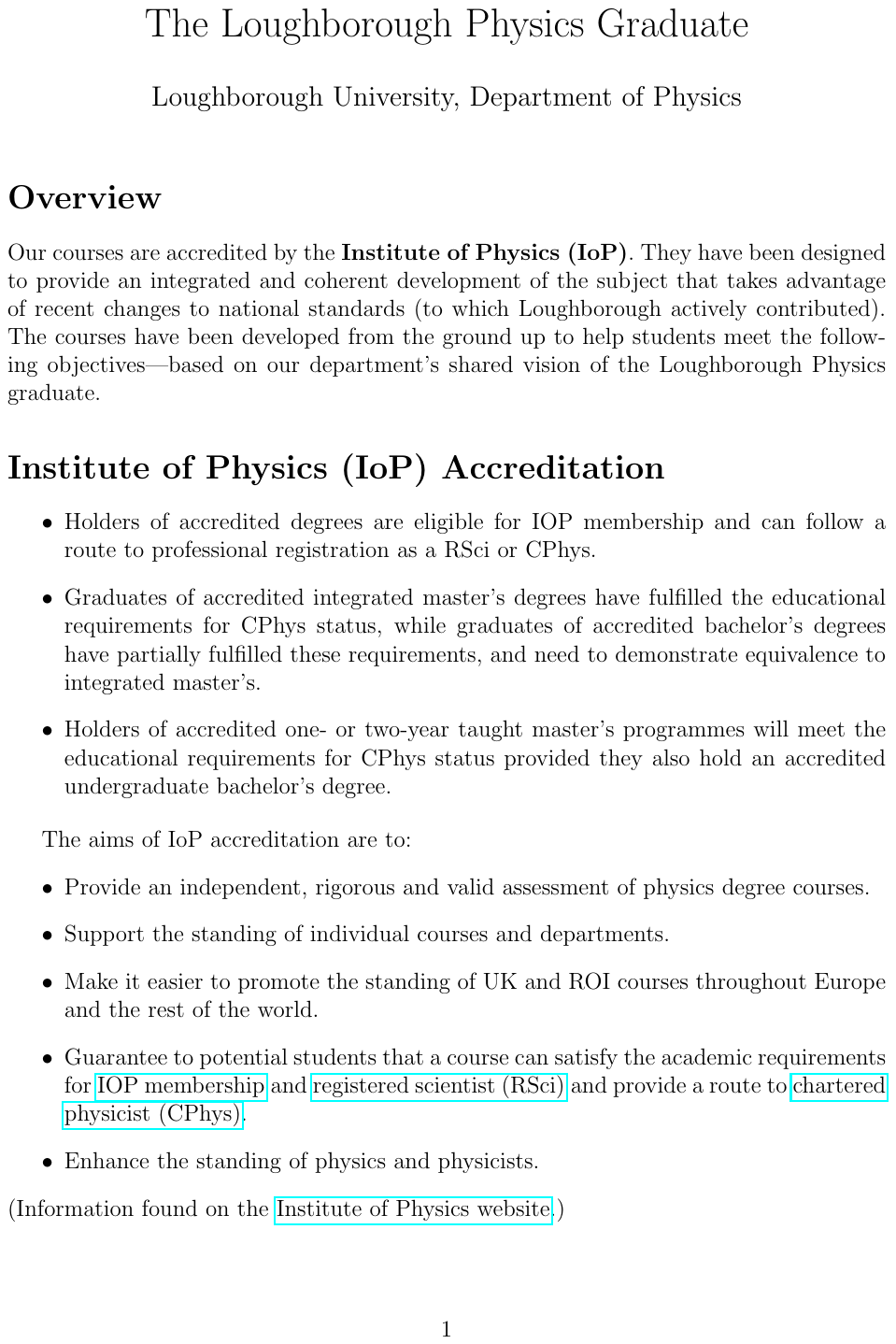}
\includepdf[pages={2},frame,scale=.95,pagecommand={}]{LoughboroughPhysicsUG}
\includepdf[pages={3},frame,scale=.95,pagecommand={}]{LoughboroughPhysicsUG}

\clearpage

\includepdf[pages={1},frame,scale=.95,pagecommand=\section{Example requirement specification\label{AppC}}]{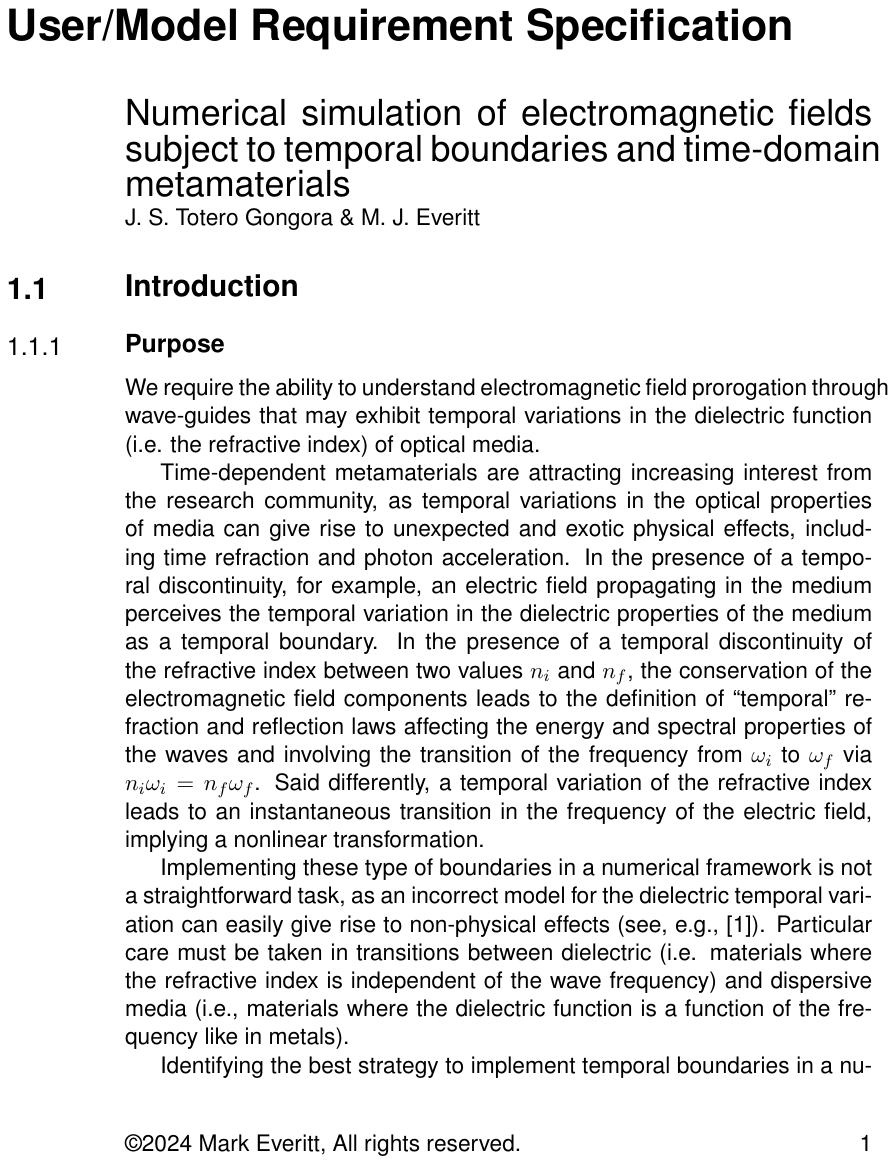}
\includepdf[pages=2,frame,scale=.95,pagecommand={}]{ModelRequiementPHA904}
\includepdf[pages=3,frame,scale=.95,pagecommand={}]{ModelRequiementPHA904}
\includepdf[pages=4,frame,scale=.95,pagecommand={}]{ModelRequiementPHA904}
\includepdf[pages=5,frame,scale=.95,pagecommand={}]{ModelRequiementPHA904}
\newpage

\includepdf[pages={2},frame,scale=.95,pagecommand=\section{Example Project Scope: Seabed Man-Made Object Detection System\label{Seabed}}]{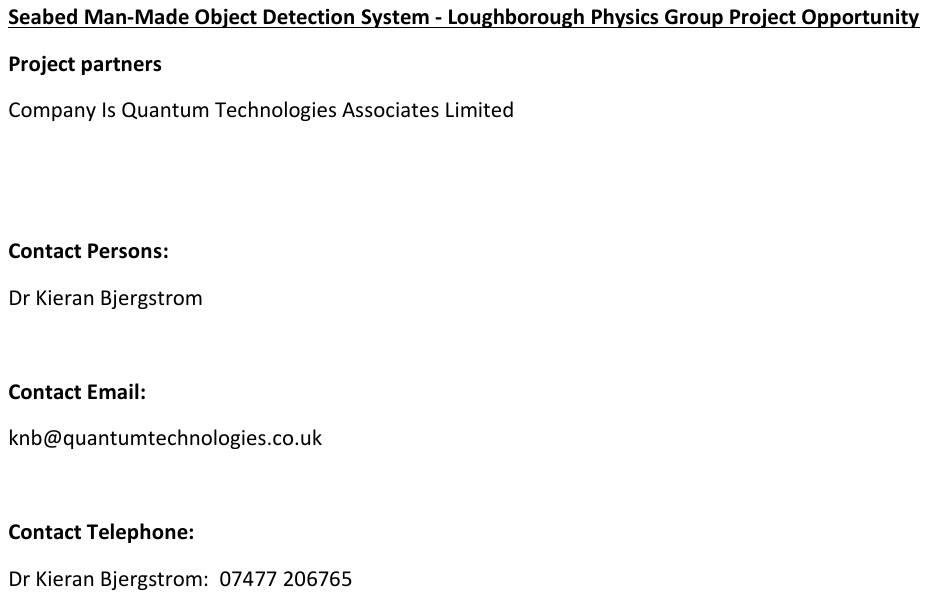}

\section{Systems engineering processes and tools}\label{toolstable}

Selected systems-engineering process and  tools that can add-value to degree-programme design in conjunction with established pedagogical tools such as Assessment Blueprints/matrices and curriculum maps.
\small
\begin{description}[style=nextline,
leftmargin=3.4cm, labelwidth=3.2cm,
font=\bfseries]

\item[Ideation \& contextualisation]
\textit{Primary SE tool(s):} Rich Picture, Root Definition, Influence Diagram, Force Field analysis.\par
\textit{Key artefact / outcome:} Shared understanding of problem context and initial mental models.

\item[Needs capture]
\textit{Primary SE tool(s):} Context Diagram; Stakeholder Influence Map; Need--Means Analysis; Viewpoint Analysis.\par
\textit{Key artefact / outcome:} Mapping of external interfaces; consolidated stakeholder-needs list.

\item[Requirements derivation]
\textit{Primary SE tool(s):} House of Quality (QFD); Holistic Requirements Model.\par
\textit{Key artefact / outcome:} Weighted stakeholder needs converted into operational, functional and non-functional requirements.

\item[Requirements baseline]
\textit{Primary SE tool(s):} Requirements Traceability Matrix (RTM).\par
\textit{Key artefact / outcome:} Bidirectional links between each requirement and future verification activity.

\item[Concept generation]
\textit{Primary SE tool(s):} Morphological Box (surprisingly powerful given its simplicity).\par
\textit{Key artefact / outcome:} Exhaustive set of curriculum-delivery concepts and trade-space definition.

\item[Concept selection]
\textit{Primary SE tool(s):} Pugh Matrix; Analytic Hierarchy Process (AHP).\par
\textit{Key artefact / outcome:} Preferred curriculum concept justified by multicriteria decision scoring.

\item[Curriculum architecture]
\textit{Primary SE tool(s):} Functional Modelling; Function--Means Analysis; Design-Structure Matrix (DSM).\par
\textit{Key artefact / outcome:} Thread-and-weave map showing dependencies and minimised prerequisite loops.

\item[Module-level risk analysis]
\textit{Primary SE tool(s):} Functional FMEA (FFMEA).\par
\textit{Key artefact / outcome:} Risk-priority numbers and mitigations for teaching, assessment and safety hazards.

\item[Detailed build]
\textit{Primary SE tool(s):} CMS, preferably a CIS with adequate capability; N$^{\mathbf{2}}$ Interface Analysis.\par
\textit{Key artefact / outcome:} Version-controlled learning materials, verified data \& resource interfaces.

\item[Component verification]
\textit{Primary SE tool(s):} Requirements Traceability Matrix column cross-check; Hypothesis testing.\par
\textit{Key artefact / outcome:} Evidence that every learning outcome and assessment item meets an actual programme (not module) level requirement (helps to avoid over-assessment that can arise from module level `optimisation').

\item[System verification]
\textit{Primary SE tool(s):} Programme-level Test Matrix (this is testing the programmes not student assessment); workload stress-test.\par
\textit{Key artefact / outcome:} Proof that assembled curriculum satisfies programme-level requirements. Hypothesis testing.

\item[Stakeholder validation]
\textit{Primary SE tool(s):} Industrial Advisory Board; Graduate focus groups; More-of/Less-of analysis; Pareto analysis.\par
\textit{Key artefact / outcome:} Confirmation that the degree meets stakeholder needs; quotations for publicity.

\item[Continuous improvement]
\textit{Primary SE tool(s):} Plan-Do-Check-Act (PDCA) cycle; annual FFMEA refresh; Is/Is not analysis; More-of/Less-of analysis; Pareto analysis.\par
\textit{Key artefact / outcome:} Updated risk register and requirement tweaks feeding the next design spiral.

\end{description}

\clearpage

\end{document}